\documentclass[%
 reprint,
 superscriptaddress,
 amsmath,amssymb,
 aps,
pra,
floatfix,
]{revtex4-2}

\usepackage{enumitem}
\usepackage{xcolor}
\usepackage{graphicx}
\usepackage{dcolumn}
\usepackage{bm}
\usepackage{hyperref}

\def\Superstaq/{\texttt{Superstaq}}
\def\qiskit/{\texttt{qiskit}}
\def\Jaqal/{\texttt{Jaqal}}

\renewcommand{\textcolor}[2]{#2}

\newcommand{\DavisCS}{Department of Computer Science, University of California, Davis, CA 95616}
\newcommand{\DavisECE}{Department of Electrical and Computer Engineering, University of California, Davis, CA 95616}
\newcommand{\DavisPhysics}{Department of Physics, University of California, Davis, CA 95616, USA}
\newcommand{\BerkeleyQNL}{Quantum Nanoelectronics Laboratory, Department of Physics, University of California at Berkeley, Berkeley, CA 94720, USA}
\newcommand{\BerkeleyComputation}{Applied Mathematics and Computational Research Division, Lawrence Berkeley National Laboratory, Berkeley, California 94720, USA}
\newcommand{\BerkeleyMaterials}{Materials Science Division, Lawrence Berkeley National Laboratory, Berkeley, California 94720, USA}
\newcommand{\ETH}{Institute for Theoretical Physics, ETH Zurich, 8093 Zurich, Switzerland}
\newcommand{\Sandia}{Sandia National Laboratories, Albuquerque, New Mexico 87123, United States}

\newcommand{\SWAP}{\operatorname{SWAP}}

\begin{document}

\preprint{APS/123-QED}

\title{\textcolor{red}{Digital Quantum} Simulation of Cavity Quantum Electrodynamics: Insights from Superconducting and Trapped Ion Quantum Testbeds}

\author{Alex H. Rubin}
\affiliation{\DavisECE}
\affiliation{\DavisPhysics}

\author{Brian Marinelli}
\affiliation{\BerkeleyQNL}
\affiliation{\BerkeleyComputation}

\author{Victoria A. Norman}
\affiliation{\DavisECE}
\affiliation{\DavisPhysics}

\author{Zainab Rizvi}
\affiliation{\DavisCS}

\author{Ashlyn D. Burch}
\affiliation{\Sandia}

\author{Ravi K. Naik}
\affiliation{\BerkeleyQNL}
\affiliation{\BerkeleyComputation}

\author{John Mark Kreikebaum}
\affiliation{\BerkeleyQNL}
\affiliation{\BerkeleyMaterials}

\author{Matthew N. H. Chow}
\affiliation{\Sandia}

\author{Daniel S. Lobser}
\affiliation{\Sandia}

\author{Melissa C. Revelle}
\affiliation{\Sandia}

\author{Christopher G. Yale}
\affiliation{\Sandia}

\author{Megan Ivory}
\affiliation{\Sandia}

\author{David I. Santiago}
\affiliation{\BerkeleyQNL}
\affiliation{\BerkeleyComputation}

\author{Christopher Spitzer}
\affiliation{\BerkeleyQNL}
\affiliation{\BerkeleyComputation}

\author{Marina Krstic-Marinkovic}
\affiliation{\ETH}

\author{Susan M. Clark}
\affiliation{\Sandia}

\author{Irfan Siddiqi}
\affiliation{\BerkeleyQNL}
\affiliation{\BerkeleyComputation}

\author{Marina Radulaski}
\affiliation{\DavisECE}

\date{\today}

\begin{abstract}
\textcolor{red}{We explore the potential for hybrid development of quantum hardware where currently available quantum computers simulate open Cavity Quantum Electrodynamical (CQED) systems for applications in optical quantum communication, simulation and computing.}
Our simulations make use of a recent quantum algorithm that maps the dynamics of a singly excited open Tavis-Cummings model containing $N$ atoms coupled to a lossy cavity.
We report the results of executing this algorithm on two noisy intermediate-scale quantum computers: a superconducting processor and a trapped ion processor, to simulate the population dynamics of an open CQED system featuring $N = 3$ atoms.
By applying technology-specific transpilation and error mitigation techniques, we minimize the impact of gate errors, noise, and decoherence in each hardware platform, obtaining results which agree closely with the exact solution of the system.
\textcolor{red}{These results can be used as a recipe for efficient and platform-specific quantum simulation of cavity-emitter systems on contemporary and future quantum computers.}
\end{abstract}

\maketitle

\section{Introduction}
Since quantum computers were first conceived of in the 1980s, a major motivation for their construction has been the possibility of efficient simulation of quantum physics \cite{Feynman1982}.
Indeed, efficient methods of implementing the unitary time evolution operator $e^{-iHt}$ are known for broad classes of physically relevant Hamiltonians $H$, requiring a quantum circuit whose size is polynomial in the degrees of freedom of $H$  \cite{patton2023polariton, doi:10.1137/060648829, Berry2007}.
Meanwhile, classical solutions of the same problem are superpolynomial at best \cite{LLoydSims}.
It is hoped that this efficiency can be used to solve physical chemistry problems related to drug development and battery manufacturing, among other areas \cite{Daley2022, PhysRevResearch.4.023019}.

Another field in which the classical intractability of large quantum simulations is problematic is the design of many-body cavity quantum electrodynamical devices.
These are systems which feature multiple quantum emitters (e.g. quantum dots, color centers, or atoms) coupled to an optical cavity or a coupled cavity array, and can form the building blocks of technologies such as quantum repeaters, simulators, and computers \cite{10.1063/5.0077045, Moody_2022}.
The use of optics provides natural advantages in certain areas of application; for example, the dominance of fiber optics in long-haul telecommunications makes CQED physics essential in designing repeater nodes for a quantum internet \cite{Kimble2008}.
Crucially, these devices must be modeled as open quantum systems in order to capture their optical emission and absorption and to understand the impact of the various forms of decoherence on their performance.
Therefore the well-studied techniques of Hamiltonian simulation are not sufficient, and we must instead apply methods which can account for the non-unitarity inherent in open quantum systems.

Fortunately, several methods for implement efficient simulations of Lindbladian dynamics have been developed in recent years, with applicability to quite general systems with at most polynomial overhead \cite{PhysRevLett.107.120501, cleve2019efficient, Childs2017, DiCandia2015, WML1, WML2}.
Given the limited sizes and coherence times of today's quantum hardware, even this modest overhead is still generally too much for practical application of these methods to technologically interesting CQED systems, a challenge that is being tackled by the development of model-specific algorithms and timely growth of quantum computational capabilities.
However, the practical aspects of executing these algorithms on specific quantum platforms need to be investigated to maximize the near term utility.

In this work, we bridge the gap between algorithm requirements and hardware capabilities by exploring methods to optimize the simulation of \textcolor{red}{an} open CQED system on two leading quantum computing platforms: one based on superconducting qubits and the other using trapped ions.
The algorithm at hand is a mapping of the singly excited Tavis-Cummings system of $N$ two-level atoms coupled to a lossy cavity \cite{Marinkovic2023} onto a quantum circuit.
By reducing the Hilbert space of each component in the system to two dimensions, this mapping reduces hardware requirements to circuit depths suitable for implementation on currently available quantum processing units (QPUs).
Through the use of optimized compilation and error mitigation strategies (such as mirror SWAPs, randomized compiling, and noiseless output extrapolation), we raise the performance of the system so that accurate simulation results can be obtained.

\section{Tavis-Cummings algorithm \label{sect:algorithm}}
\begin{figure}[htbp]
    \centering
    \includegraphics[width=\columnwidth]{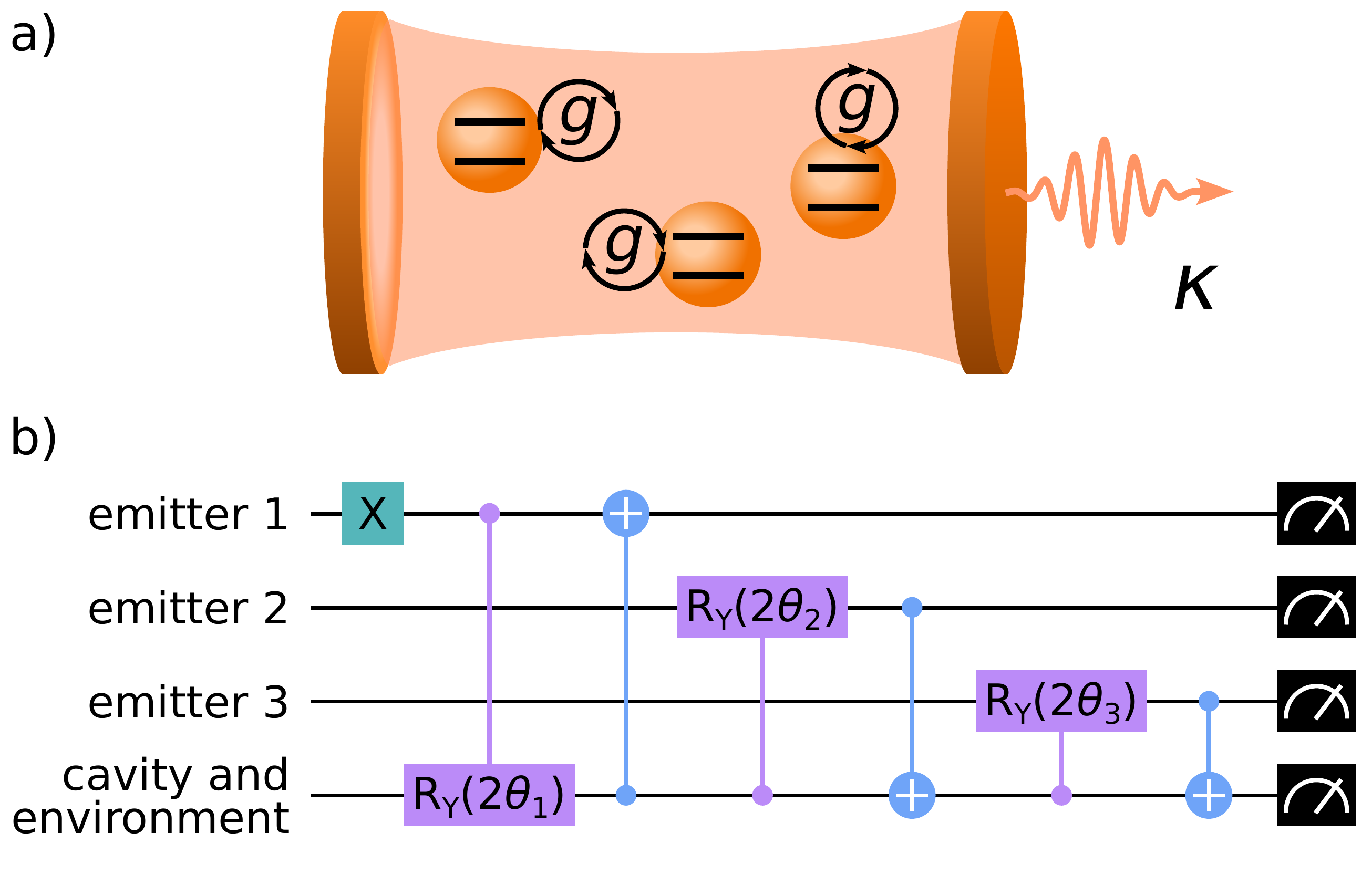}
    \caption{(a) Diagram of an open Tavis-Cummings CQED system consisting of three two-level quantum emitters coupled to a lossy optical cavity mode. (b) The Q-MARINA circuit for an $N=3$ emitter Tavis-Cummings system used in the simulations presented in this work. The initial state is emitter 1 in the excited state, and all other parts of the system are in their ground states.}
    \label{fig:qmarina}
\end{figure}

The Tavis-Cummings model \cite{PhysRev.170.379} describes $N$ two-level quantum emitters coupled to a single optical mode (see Fig.~\ref{fig:qmarina}(a)) with the Hamiltonian
\begin{equation}
    H_{\text{TC}} = \omega_c a^\dag a + \sum_{i=1}^N \left\{ \frac{1}{2} \omega_i \sigma^z_i + g_i (\sigma^+_i a + a^\dag \sigma^-_i) \right\},
\end{equation}
where $\omega_c$, $\omega_i$ are the frequencies of the cavity mode and the $i^\text{th}$ emitter, respectively; $\sigma^z_i$, $\sigma^\pm_i$ are the Pauli $Z$ and raising/lowering operators for the $i^\text{th}$ emitter, respectively; $a$ ($a^\dagger$) is the annihilation (creation) operator for the cavity mode; and $g_i$ is the coupling of the $i^\text{th}$ emitter to the cavity mode. Here and throughout this discussion we use natural units ($\hbar=1$).

Including Markovian interactions with the environment in the form of optical loss from the cavity mode, time evolution of the density matrix $\rho$ for the system obeys a master equation which takes the following Lindblad form
\begin{equation}
    \dot{\rho} = \mathcal{L}(\rho) \equiv -i[H_{\text{TC}}, \rho] + \frac{\kappa}{2} \mathcal{D}_a(\rho),
    \label{eq:master-eq}
\end{equation}
where $\mathcal{D}_a(\rho) = 2a\rho a^\dagger - \{a^\dagger a, \rho\}$ is the dissipator generated by $a$, representing loss from the cavity at rate $\kappa$.

Note that this master equation neglects blackbody excitation of the cavity, because the average thermal occupation of modes in the technologically relevant wavelength range ($\sim 1$ $\mu$m) are extremely low at room temperature and below.
For simplicity, we also neglect other decoherence processes such as nonradiative relaxation and phase decay of the emitters.
As described in the foregoing section, numerically solving \eqref{eq:master-eq} for a particular value of time $t$ as
\begin{equation}
    \rho(t) = e^{\mathcal{L}t}(\rho(t=0))
\end{equation}
using classical methods incurs $\mathcal{O}(2^N)$ memory and runtime costs in general.
We therefore seek efficient quantum algorithms for this purpose.

\textcolor{red}{The recently developed Q-MARINA algorithm maps this special case of the open Tavis-Cummings system onto a quantum circuit with depth $2N+1$ gates acting on $N+1$ qubits \cite{Marinkovic2023}, achieving $\mathcal{O}(N)$ space and time cost.
It is important to note that this mapping works only in the single-excitation regime, which is efficiently solvable by classical methods as well.}
\textcolor{red}{Nevertheless, its low resource requirements make this method practical with current quantum processors, and therefore presents an opportunity for exploring the feasibility of achieving accurate simulations of CQED physics on today's noisy intermediate-scale quantum (NISQ) hardware.}
\textcolor{red}{Because we consider only a single excitation, the combined population of the cavity/environment can be represented by a single qubit.}
\textcolor{red}{By bringing the environment into the simulation explicitly, the dissipative dynamics of the system can be implemented with the unitary operations available on a quantum processor.
The $N$ two-level emitter populations are likewise mapped onto $N$ qubits.}

\textcolor{red}{Starting with an initial state where one emitter is excited and the rest of the system is in the ground state, the simulation circuit calculates the expectation value of the populations of each emitter ($\langle \sigma_i^\dagger \sigma_i \rangle$) and the cavity/environment at some arbitrary time $t$ in the future.}
\textcolor{red}{The population transfer between the cavity/environment qubit and each emitter qubit is computed with a controlled-Y rotation followed by a CNOT acting on those two qubits.}
\textcolor{red}{The angle of the controlled-Y rotation is a function of the Tavis-Cummings system parameters ($g$, $\kappa$), the propagation time $t$, and the rotation angles of each cavity-emitter interaction calculated previously in the circuit: $\theta_i = \theta_{i}(g, \kappa, t, \theta_{1},\dots,\theta_{i-1})$.}
\textcolor{red}{The overall structure of a Q-MARINA circuit (see Fig.~\ref{fig:qmarina}(b)) is as follows:
\begin{enumerate}[nosep]
    \item Prepare the initial state with an X gate on one emitter qubit, representing an excitation.
    \item The excited emitter qubit interacts with the cavity/environment qubit through a controlled-Y rotation and a CNOT. 
    \item Each other emitter qubit interacts with the cavity/environment qubit (via a controlled-Y rotation and a CNOT) in turn.
\end{enumerate}}

\textcolor{red}{The structure of this algorithm as a direct mapping between physical system and quantum circuit stands in contrast to other algorithms for simulating general open quantum systems (\cite{PhysRevLett.107.120501, cleve2019efficient, Childs2017, DiCandia2015, WML1, WML2}) which rely on either a Trotter decomposition or the use of carefully chosen measurements to implement the dissipative dynamics.
As a result, these methods have a tradeoff between the number of Trotter steps or measurements, and the final accuracy.
Q-MARINA directly implements the exact dynamics by including the environment in the simulation, so its accuracy boils down to the errors inherent in the quantum processor it is implemented on.}

\textcolor{red}{We use the Q-MARINA algorithm to simulate an $N=3$ open Tavis-Cummings system on two NISQ devices: a trapped ion processor, and a superconducting processor.}
\textcolor{red}{We benchmark the results from the two quantum processors against the numerical solution to the master equation (\ref{eq:master-eq}) obtained using QuTiP (Quantum Toolbox in Python), a software package for classical simulations of quantum systems \cite{qutip}.}

\section{\textcolor{red}{Quantum error mitigation}}

\textcolor{red}{Since the 1990s, work by Shor and others has shown that it's possible in principle to detect and correct errors in a quantum processor provided the physical error rates remain below certain thresholds \cite{PhysRevA.54.1098, doi:10.1126/science.279.5349.342}.
Quantum computing hardware has made great strides in the decades since then, and recently some groups have demonstrated the first logical qubits operating below threshold \cite{acharya2024quantumerrorcorrectionsurface, paetznick2024}.
However these error thresholds remain quite challenging to meet, and systems featuring error-corrected qubits are still in their infancy.}

\textcolor{red}{In light of these challenges, researchers have developed a suite of techniques known as quantum error mitigation (QEM) that can partially compensate for errors without requiring the significant quantum resources needed for full error correction \cite{RevModPhys.95.045005}.
QEM techniques typically take advantage of features of a quantum processor's noise profile, implementation details of its gates, or structure implicit to a given quantum circuit, and use this to guide a circuit's compilation or postprocessing of measurement results in order to minimize the impact of noise.
Therefore QEM is most effectively used by taking into account the details of the circuit and hardware platform.
For example, randomized compiling (RC) acts to convert coherent errors into stochastic ones by compiling many different (but logically equivalent) versions of a circuit; platforms in which coherent errors dominate will thus be best suited to using RC.
}

\subsection{\textcolor{red}{Model-specific postselection and averaging} \label{sec:postselect}}

\textcolor{red}{
A particularly simple form of error mitigation we employed in our simulations is the post-processing applied to the raw measurement results, which is guided by the physics of our simulations.
First, we postselected our data by excluding any measurements which violated conservation of energy in the simulated system.
As previously described, the initial state of the system represents a single excitation in one of the emitters; since the environment is included in the simulation, in the absence of noise the total population in the system should remain fixed.
We therefore discard any measurement results which fall outside the single-excitation subspace, as they represent errors.
Unlike other postselection schemes where a fixed fraction of events must be discarded due to inherent probabilistic effects, our method's discard rate depends only on the quality of the experimental implementation.
This means that as hardware improves, the postselection overhead could approach zero, allowing the technique to scale to larger simulations.
In our simulations the fraction of discarded shots ranged from 8-29\% depending on the specific implementation.
}

\textcolor{red}{As a second postprocessing step, we note that the simulated populations of emitters two and three should be the same at all time steps since they are treated identically in the master equation (\ref{eq:master-eq}) and have the same initial state (the ground state).
Any discrepancy between their populations represents the effect of noise, so after postselection we average these two emitters' populations and consider that the population of both.
This is essentially equivalent to doubling the number of shots on this portion of the simulation, reducing sampling error by a factor of $1/\sqrt{2}$, or $1/\sqrt{N-1}$ for systems with $N$ emitters.
All data presented in this work has been postselected and averaged in the manner just described unless otherwise stated.}

\section{Implementation on a trapped ion quantum testbed}

\subsection{The QSCOUT trapped ion quantum processor}
The first implementation of the algorithm took place on a trapped ion quantum processor, the Quantum Scientific Computing Open User Testbed (QSCOUT) at Sandia National Laboratories \cite{qscout_manual}.
The platform consists of four ytterbium $(^{171}$Yb$^{+})$ ion qubits in a linear trap.
All-to-all connectivity is provided by coupling to the vibrational modes of the ion chain. 
This high connectivity lends itself well to the algorithm, which requires one-to-all interactions.

On the hardware, all Raman transitions are realized with combinations of individual addressing beams and a global beam counter-propagating to the individual addressing beams' direction.
The single-qubit gates are realized by placing both tones of the Raman transition on an individual beam (co-propagating), while the two-qubit gates are realized by placing the necessary tones across both the individual and global beams (counter-propagating).
We use QSCOUT's single qubit X gate and two-qubit arbitrary-angle M{\o}lmer–S{\o}rensen gate, which can be expressed on the hardware in either the XX basis (MS$_\text{XX}$), or a ZZ basis variation.
While the bare MS interaction is XX-type, both versions require a series of wrapper single-qubit gates in order to mitigate phase instabilities when mixing gates of differing propagation~\cite{lee2005} and to reduce phase-dependent crosstalk~\cite{chow2024}.
The fidelities of the MS and single-qubit gates are given in Appendix~\ref{app:qscout-fidelity}.
A high-level programming language, \Jaqal/, is provided for interfacing with the hardware.

We executed the Q-MARINA algorithm on this platform to simulate a CQED system with the following parameters: $N=3$, \textcolor{red}{$g=4$ GHz, $\kappa=2$ GHz}; the system was evolved in time from $t=0$ ns to $t_\text{max}=3$ ns.
\textcolor{red}{We benchmark our quantum simulations against the results of a classical simulation obtained with the python package QuTiP.}

\subsection{Circuit compilations}

\begin{figure*}[htbp]
    \centering
    \includegraphics[width=\textwidth]{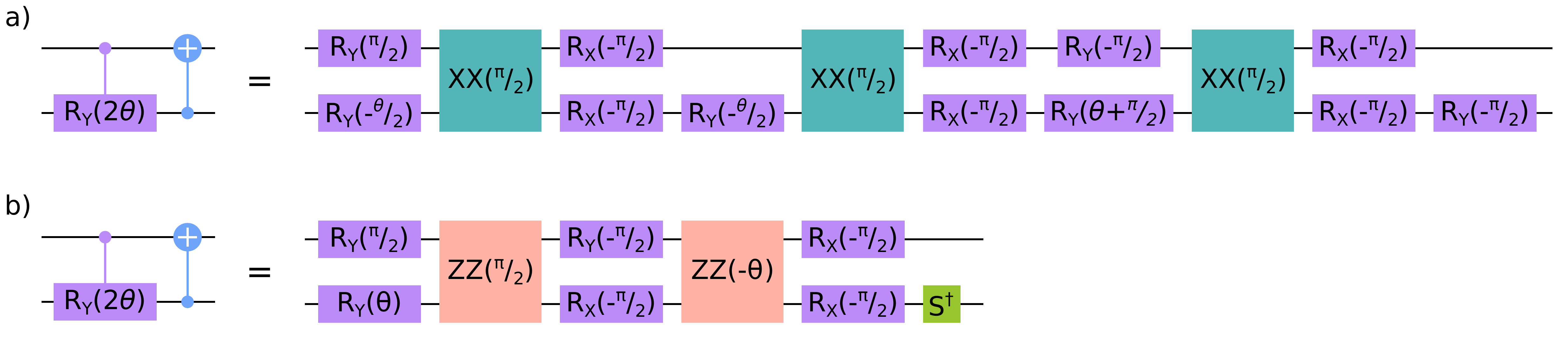}
    \caption{
    \textcolor{red}{Different compilations of the gates implementing an interaction between the emitter qubit and the environment qubit, using QSCOUT native gates. (a) The manually compiled version  derived as described in the main text. (b) An example of a compilation using ZZ entangling gates, produced by \Superstaq/. Note that while the compiled circuit always contains a ZZ$(\pi/2)$ and ZZ$(-\theta)$, the optimized single-qubit wrapper gates vary depending on the value of $\theta$, so this exact compilation was not used in every time step of the simulation, and is presented only as an example.}
    }
    \label{fig:qscout-compilations}
\end{figure*}

We first manually compiled the algorithm to QSCOUT's native gates by hand and implemented it in the programming language \Jaqal/.
Each controlled-R$_\text{Y}$ operation was compiled into a pair of CNOT gates and unconditional R$_\text{Y}$ rotations \cite{MikeAndIke} and further decomposed each CNOT into MS$_\text{XX}$ and single-qubit rotations \cite{Maslov_2017}, yielding the circuit seem in Fig.~\ref{fig:qscout-compilations}(a).

\begin{figure}[htbp]
    \centering
    \includegraphics[width=\columnwidth]{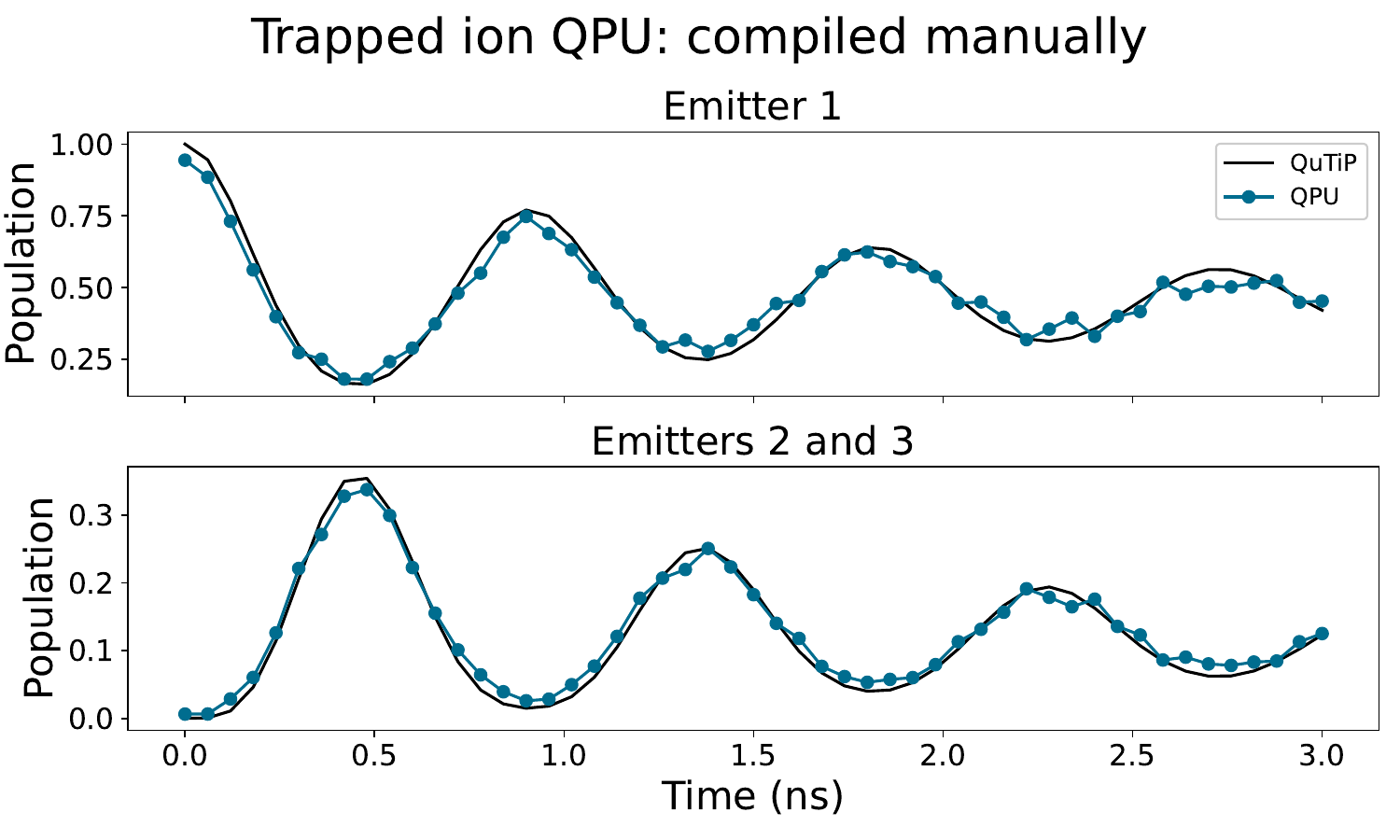}
    \caption{
    \textcolor{red}{Results from the QSCOUT trapped ion QPU using the manually compiled version of the algorithm.
    Each data point represents 2,000 shots; approximately 27\% of shots were discarded in postselection using the technique described in Sec.~\ref{sec:postselect}.
    Wall-clock runtime of the simulation was about eight minutes.
    Black line indicates the exact solution of the system.}
    }
    \label{fig:qscout-jaqal-raw}
\end{figure}

While preserving the $\mathcal{O}(N)$ circuit depth, this manual compilation increased the overall number of two-qubit gates in each circuit from $2N$ to $3N$, an overhead which would be expected to impact performance given that two-qubit gates are typically the most challenging operations in a quantum computer.
The results of this hand-compiled version of the simulation can be seen in Fig.~\ref{fig:qscout-jaqal-raw}.

\begin{figure}[htbp]
    \centering
    \includegraphics[width=\columnwidth]{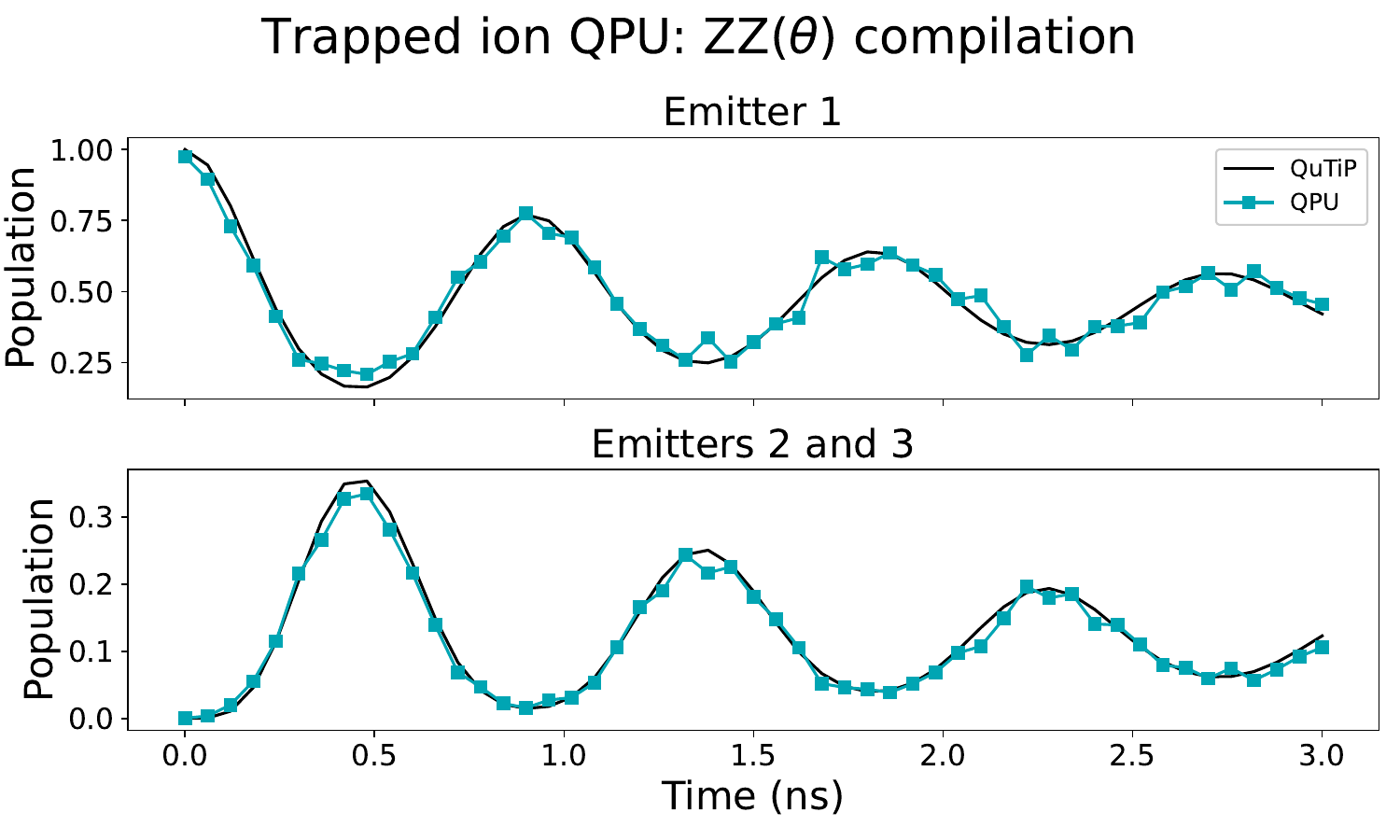}
    \caption{
    \textcolor{red}{Results from the QSCOUT trapped ion QPU using the version of the simulation circuit compiled with ZZ($\theta$) entangling gates by \Superstaq/.
    Each data point represents a simulation with 2,000 shots; approximately 10\% of shots were discarded in postselection using the technique described in Sec.~\ref{sec:postselect}.
    Wall-clock runtime of the simulation was about three minutes.
    Black line indicates the exact solution of the system. 
    }
    }
    \label{fig:qscout-superstaq}
\end{figure}

We improved upon the use of fixed MS entangling gates in the hand-compiled version of our algorithm by switching to ZZ($\theta$) entangling gates.
This was done using \Superstaq/, an optimizing compiler for quantum programs \cite{campbell2023superstaq, yale2024}.
Starting with an implementation of our algorithm written in \qiskit/ (see Fig.~\ref{fig:qmarina}(b)) \Superstaq/ compiled it to QSCOUT's native gates, with ZZ($\theta$) as the two-qubit entangling gate.
This compilation method produced native circuits containing $2N$ two-qubit gates (see Fig.~\ref{fig:qscout-compilations}(b)), matching the two-qubit gate complexity of the original abstract circuit (Fig.~\ref{fig:qmarina}(b)).
Results of executing this \Superstaq/-compiled version of Q-MARINA are presented in Fig.~\ref{fig:qscout-superstaq}.

\subsection{Error mitigation: SWAP mirroring}
\begin{figure}[htbp]
    \centering
    \includegraphics[width=\columnwidth]{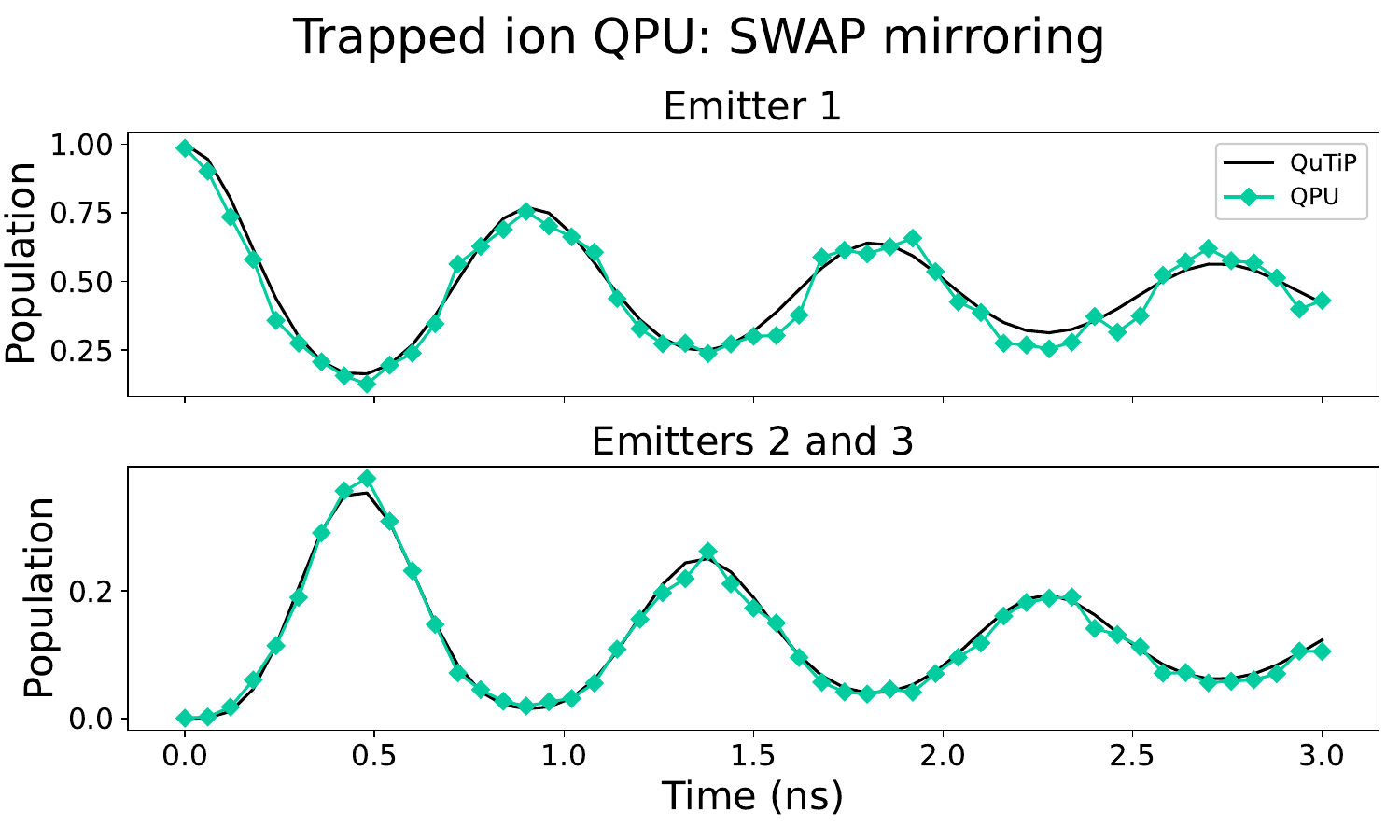}
    \caption{
    \textcolor{red}{Results from the QSCOUT trapped ion QPU
    using the algorithm compiled with ZZ($\theta$) entangling gates and SWAP mirroring by \Superstaq/.
    Each data point represents a simulation with 2,000 shots; approximately 10\% of shots were discarded in postselection using the technique described in Sec.~\ref{sec:postselect}.
    Wall-clock runtime of the simulation was about three minutes.
    Black line indicates the exact solution of the system.}
    }
    \label{fig:qscout-mirror-swaps}
\end{figure}

To attempt to further improve the simulation accuracy, we applied an error mitigation technique known as SWAP mirroring \cite{campbell2023superstaq}, which is implemented as a feature within the \Superstaq/ quantum compiler.
SWAP mirroring is based on the observation that there are instances when appending two SWAP gates to an arbitrary two-qubit unitary can allow it to be compiled to a more efficient gate sequence than the original unitary itself, since the final SWAP can be achieved by virtual qubit relabeling.
\textcolor{red}{The \Superstaq/ compiler compares versions of the circuit compiled with and without appended SWAPs, and selects the implementation which minimizes the total MS rotation angle.
Smaller MS gate angles require lower laser amplitude to implement, and are thus subject to reduced amplitude noise.
The total MS gate angle across the simulation was reduced by about 2\%; the MS gates which were improved by SWAP mirroring had their rotation angles reduced by about 17\% on average.}
The results of executing our algorithm compiled by \Superstaq/ with ZZ($\theta$) entangling gates and SWAP mirroring is seen in Fig.~\ref{fig:qscout-mirror-swaps}.

\subsection{Error mitigation: randomized compiling \label{sec:rc}}
\begin{figure}[htbp]
    \centering
    \includegraphics[width=\columnwidth]{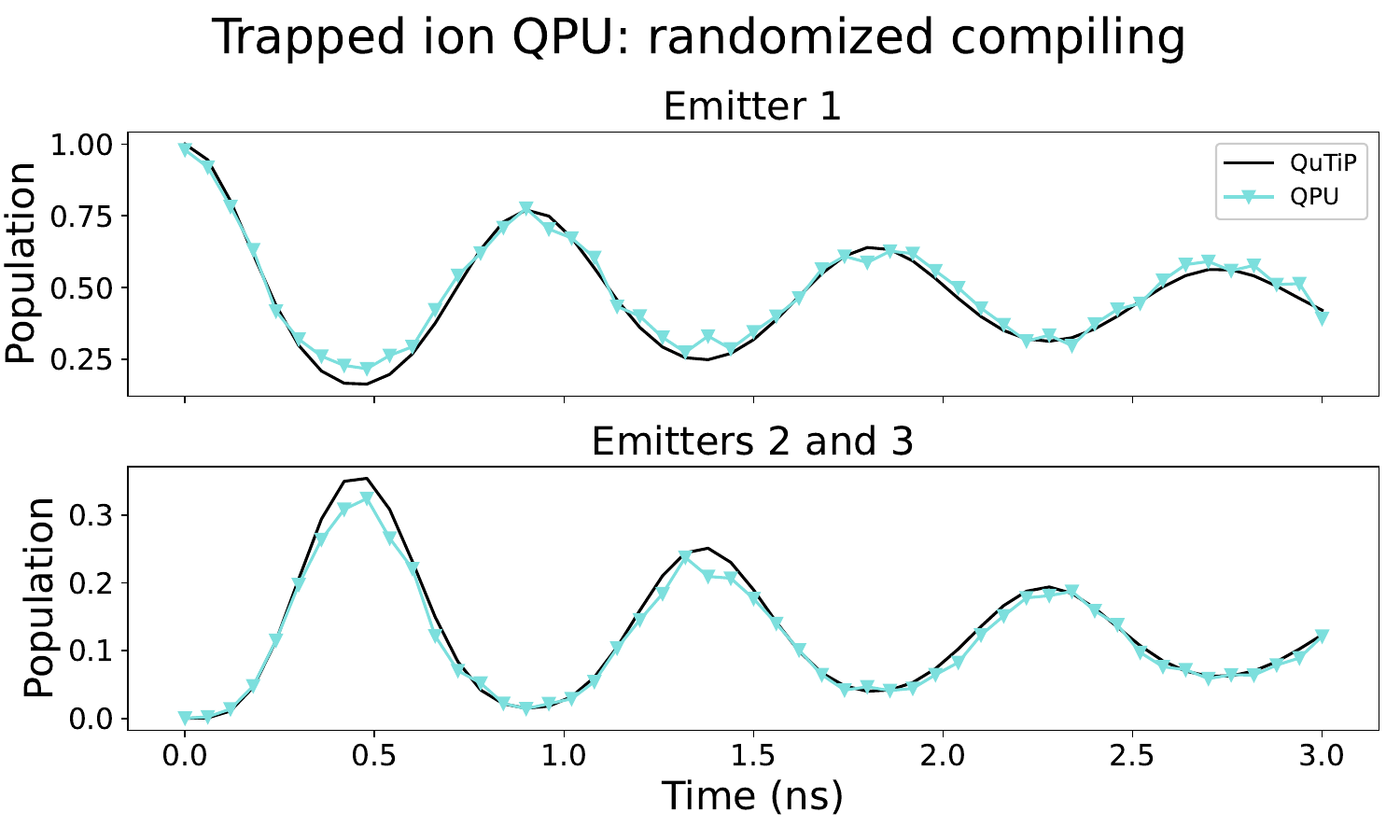}
    \caption{
    \textcolor{red}{
    Results from the QSCOUT trapped ion QPU
    using the algorithm compiled with ZZ($\theta$) entangling gates, averaged over 10 randomized compilations.
    Each data point represents a simulation with 2,000 shots; approximately 15\% of shots were discarded in postselection using the technique described in Sec.~\ref{sec:postselect}.
    Wall-clock runtime of the simulation was about one hour.
    The black line indicates the exact solution of the system.
    }
    }
    \label{fig:qscout-rc}
\end{figure}

We also applied error mitigation by using randomized compiling \cite{PhysRevX.11.041039, PhysRevA.94.052325}.
This technique adds randomly selected single-qubit gates around each two-qubit gate cycle, chosen in such a way that the overall logical operation of the circuit is unchanged.
Many equivalent randomized versions of the circuit are produced, each is run for some number of shots, and an average is taken over all the results.
The effect of this randomization is to convert coherent and non-Markovian errors into uncorrelated stochastic errors; a larger number of randomizations more strongly tailors the noise.
Since coherent errors accumulate quadratically in circuit depth and stochastic errors accumulate linearly, randomized compiling provides an advantage by mitigating the growth of errors.
The results of the simulation compiled with RC into 10 randomizations are presented in Fig.~\ref{fig:qscout-rc}.

\subsection{Analysis \label{sect:hellinger}}
The behaviors of the cavity and emitter populations over time are fundamental to characterizing a CQED system's behavior.
Accurately modeling these dynamics can not only shed light on the basic physics involved (for example, Rabi oscillations, decoherence, and in some cases, collective phenomena such as super- and subradiance) but is also of central importance in designing such a system for technological applications – for instance, optimizing quantum state transduction in a CQED-based quantum repeater node.
\textcolor{red}{To quantitatively assess the accuracy of our quantum simulations we employ the Hellinger distance, a metric of the difference between two probability distributions.
At each time step, the simulation yields the populations of the constituent parts of the system, which sum to one, so the Hellinger distance between the simulated distribution and the exact distribution at time $t$ is}
\begin{equation}
    H(t) = \frac{1}{\sqrt{2}}\sqrt{\sum_{i} \left(\sqrt{s_i(t)} - \sqrt{e_i(t)}\right)^2}.
\end{equation}
\textcolor{red}{where $s_j[t]$ and $e_j[t]$ are the simulated and exact probabilities for state $j$ at time $t$.
We then take the mean of $H(t)$ across the full timespan of the simulation:}
\begin{equation}
    H = \frac{1}{t_\text{max}}\sum_{t=0}^{t_\text{max}} H(t),
    \label{eq:hellinger}
\end{equation}
\textcolor{red}{which we refer to as Mean Hellinger Distance (MHD).
This can be taken as a mean of the error of the entire simulation; a Mean Hellinger Distance $H=0$ would indicate perfect agreement between the results of the quantum processor and the exact numerical solution.}

\begin{figure}[htbp]
    \centering
    \includegraphics[width=\columnwidth]{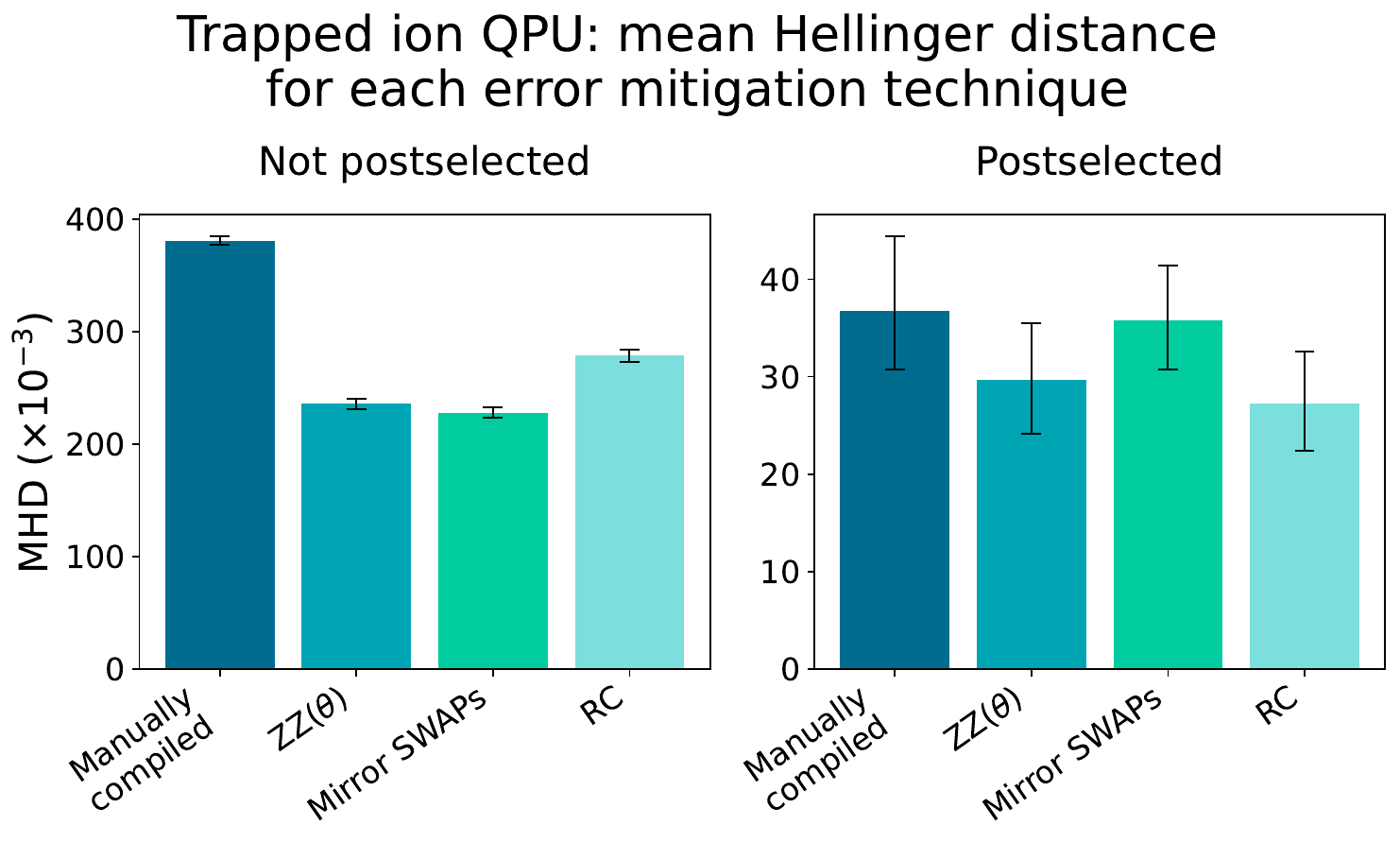}
    \caption{
    \textcolor{red}{
    Mean Hellinger Distance (see Eq.~(\ref{eq:hellinger})) between the simulated distribution and the exact numerical solution obtained with QuTiP for each version of the algorithm executed on the QSCOUT trapped ion QPU, with and without postselection applied.
    Each simulation was executed with 51 timesteps, with 2,000 shots per timestep.
    Error bars represent 95\% confidence intervals computed via bootstrap resampling with 1,000 replicates.
    }
    }
    \label{fig:qscout-hellinger}
\end{figure}

\textcolor{red}{In Fig.~\ref{fig:qscout-hellinger}, we compare the MHD for the four versions of the simulation run on the QSCOUT trapped ion processor.
When postselection is not applied, all three of the error mitigation techniques (ZZ($\theta$) compilation, mirror SWAP, and RC) all provide a similar level of reduction in MHD relative to the manually compiled version of the simulation with fixed MS gates, which performs the worst.
The randomized compiling version of the algorithm has a somewhat higher MHD than either the ZZ($\theta$) or mirror SWAP versions, likely due to the deeper circuits produced by RC.
While ion traps benefit from high gate fidelities and coherence times, their slower gate speeds mean deeper circuits take longer to execute, during which time experimental parameters like laser intensities and trap fields can drift, potentially introducing cumulative errors that are difficult to mitigate through randomization techniques.}

\textcolor{red}{The most significant improvement in MHD is provided by postselection itself, which reduces the simulation error by as much as a factor of 10.
When postselection is applied, all four versions of the algorithm exhibit roughly comparable performance.
This likely indicates that the errors affecting the simulation are predominantly stochastic, as the three error mitigation techniques act by either reducing coherent errors (ZZ($\theta$) and mirror SWAP) or tailoring them into stochastic errors for better averaging (RC).}

\begin{figure}[htbp]
    \centering
    \includegraphics[width=\columnwidth]{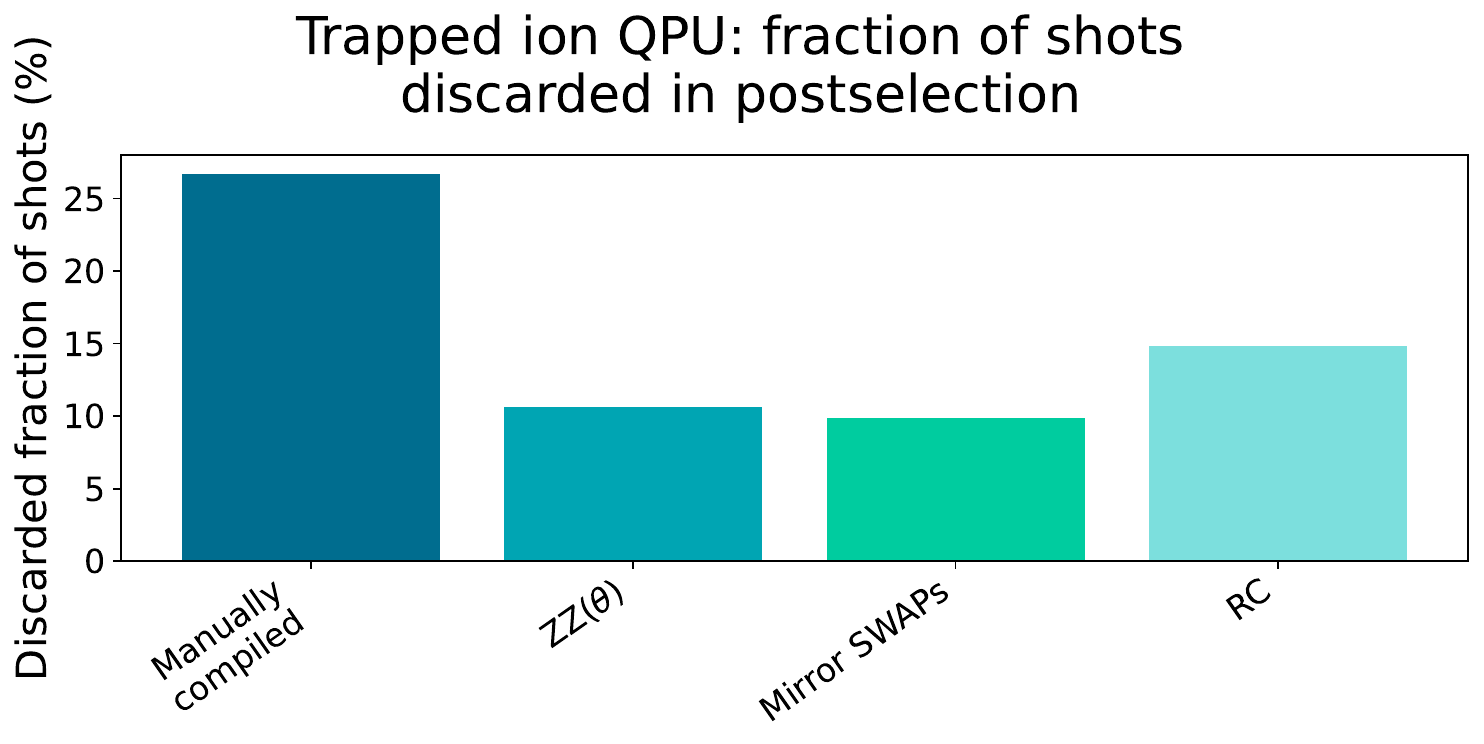}
    \caption{
    \textcolor{red}{
    Fraction of simulation shots discarded by postselection (see Sec.~\ref{sec:postselect}) for each version of the algorithm executed on the QSCOUT trapped ion QPU.
    }
    }
    \label{fig:qscout-discarded}
\end{figure}

\textcolor{red}{The effect of the error mitigation strategies can also be seen in the fraction of shots that were discarded by the postselection technique (Fig.~\ref{fig:qscout-discarded}).
As the accuracy of the simulation improves, fewer shots of the simulation fall outside the single-excitation subspace and are excluded by postselection.
These results broadly follow the trend of the non-postselected data in Fig.~\ref{fig:qscout-hellinger}.
}

\begin{figure}[htbp]
    \centering
    \includegraphics[width=\columnwidth]{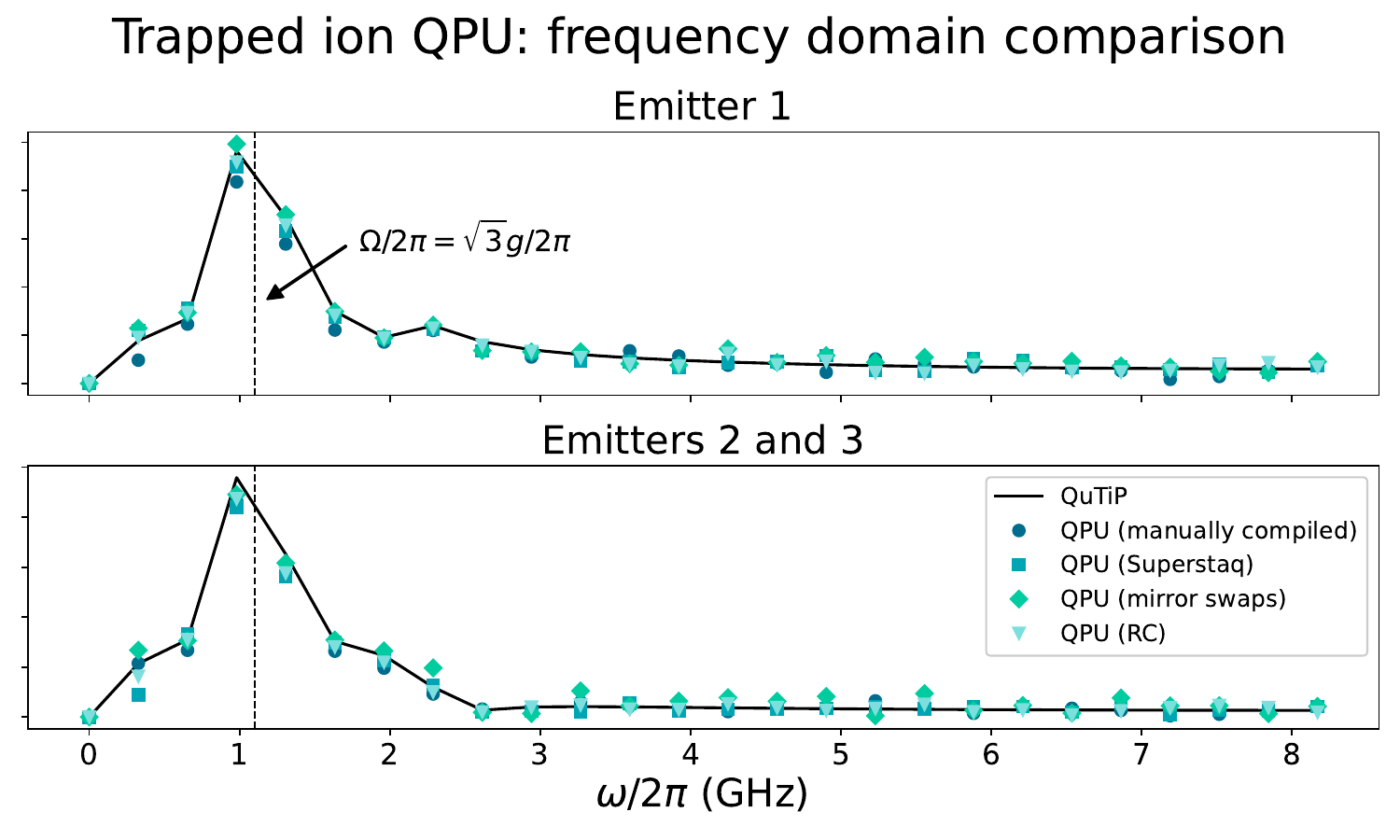}
    \caption{
    Fourier transforms of the simulation results from the QSCOUT trapped ion quantum processor.
    The mean population has been subtracted, to eliminate the zero-frequency component.
    Each data point represents a simulation with 2,000 shots.
    The data has been postselected as described in Sec.~\ref{sec:postselect}.
    The dashed line indicates the theoretical Rabi frequency for the simulated system.
    The black line indicates the Fourier transform of the numerical solution obtained with QuTiP.
    }
    \label{fig:qscout-fourier}
\end{figure}

We also examined the accuracy of the simulations in frequency space.
A central feature of a strongly coupled CQED system's behavior is Rabi oscillations: exchange of excitations between the emitters and the cavity.
Understanding the rate of Rabi oscillations is crucial for engineering control pulses needed to operate the system as a quantum networking device or component of a quantum computer.
To assess the accuracy of the simulated Rabi oscillations, we compared the Fourier transforms of the simulation results with those of the numerical solution.
These results for QSCOUT are presented in Fig.~\ref{fig:qscout-fourier} for all three versions of the simulation.
Since the simulated system consists of identical emitters resonant with the cavity, the theoretical Rabi rate is simply $\Omega = \sqrt{N}g$.
\textcolor{red}{We find that the quantum simulations accurately obtain the Rabi rate of the system, even without optimized compilation techniques or error mitigation, which is partly a reflection of the robustness of Fourier analysis to noise. Errors are largely uncorrelated between time steps in the simulation, so they appear as low-amplitude white noise in the frequency domain.
This allows the fundamental Rabi frequency to remain clearly identifiable in the Fourier transform despite the presence of amplitude noise in the time-domain data.}

\section{Implementation on a superconducting quantum testbed}

\subsection{The AQT superconducting quantum processor}
To test our algorithm on a superconducting system, we used a processor built and operated by the Advanced Quantum Testbed (AQT) at Lawrence Berkeley National Laboratory.
We use four of the superconducting transmon qubits on the device, arranged in a linear chain, such that each qubit can interact with one or two adjacent qubits.
Arbitrary single qubit rotations can be performed with resonant Rabi driving.
The native two-qubit gate is a controlled phase (CZ) gate described in Ref.~\cite{Mitchell2021}.
\textcolor{red}{The fidelities for the CZ and single-qubit gates on this device are given in \cite[Supplementary Note 4]{Sun2024}.}

As before, we simulate an open Tavis-Cummings system with parameters $N=3$, $g=4$ GHz, $\kappa=2$ GHz, and the system was evolved in time from $t=0$ ns to $t_\text{max}=3$ ns.
All results from the AQT processor were postselected according to the procedure described in Sec.~\ref{sec:postselect}; the fraction of shots discarded in postselection was about 8-29\% depending on the implementation.

\subsection{Circuit compilations}

\begin{figure}[htbp]
    \centering
    \includegraphics[width=\columnwidth]{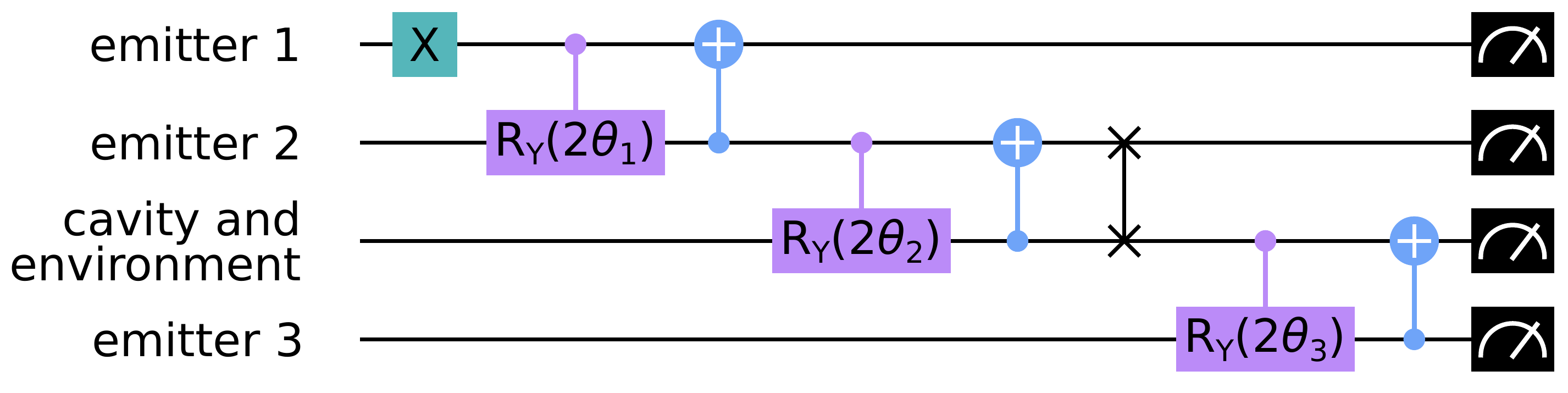}
    \caption{The modified simulation circuit executed on AQT's 4-qubit superconducting processor, showing the use of the SWAP to shuttle the cavity/environment qubit along the chain.
    The qubit labels indicate the role of the qubits at the end of the circuit, after the $\SWAP$ has interchanged the emitter 2 and cavity/environment qubits.
    }
    \label{fig:aqt-qmarina}
\end{figure}

In order to adapt the simulation algorithm's one-to-all interaction topology to the connectivity of the linear AQT processor, we added a $\SWAP$ gate to shuttle the cavity/environment qubit along the chain, allowing it to interact with each of the emitter qubits (see Fig.~\ref{fig:aqt-qmarina}).
This is analogous to the ``star-to-line" routing optimization implemented by Superstaq \cite{campbell2023superstaq}.
In principle, this could extend to any length of linear qubit chain with only $\mathcal{O}(N)$ $\SWAP$ gates, allowing for the simulation of larger systems while preserving the linear circuit depth and runtime.
However it is worth noting that, in practice, even a constant-factor increase in circuit depth can markedly affect the performance.

\begin{figure}[htbp]
    \centering
    \includegraphics[width=\columnwidth]{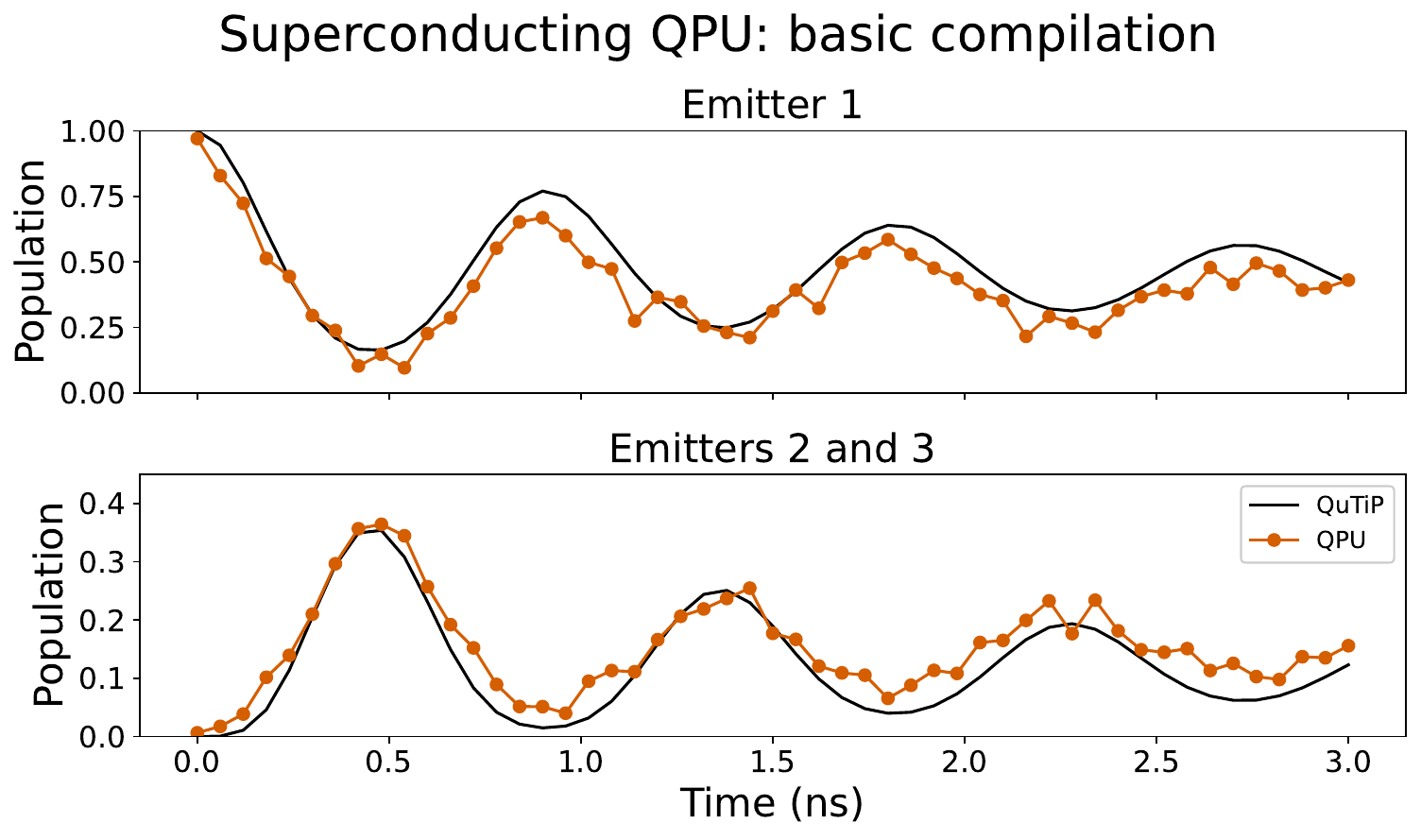}
    \caption{
    \textcolor{red}{
    Results from the AQT superconducting QPU with no error mitigation compiled into the circuit.
    Each data point represents a simulation with 20,000 shots; approximately 29\% of shots were discarded in postselection using the technique described in Sec.~\ref{sec:postselect}.
    Wall-clock runtime of the simulation was about eight minutes.
    The black line indicates the exact solution of the system.
    }
    }
    \label{fig:aqt-raw}
\end{figure}

We initially executed a version of the algorithm compiled to AQT's native gates without error mitigation.
The result of running this version of the circuit (postselected as described in Sec.~\ref{sec:postselect}) are presented in Fig.~\ref{fig:aqt-raw}, already shows clearly visible Rabi oscillations and rough agreement with the numerical solution.

\subsection{Error mitigation: randomized compiling (RC)}
\begin{figure}[htbp]
    \centering
    \includegraphics[width=\columnwidth]{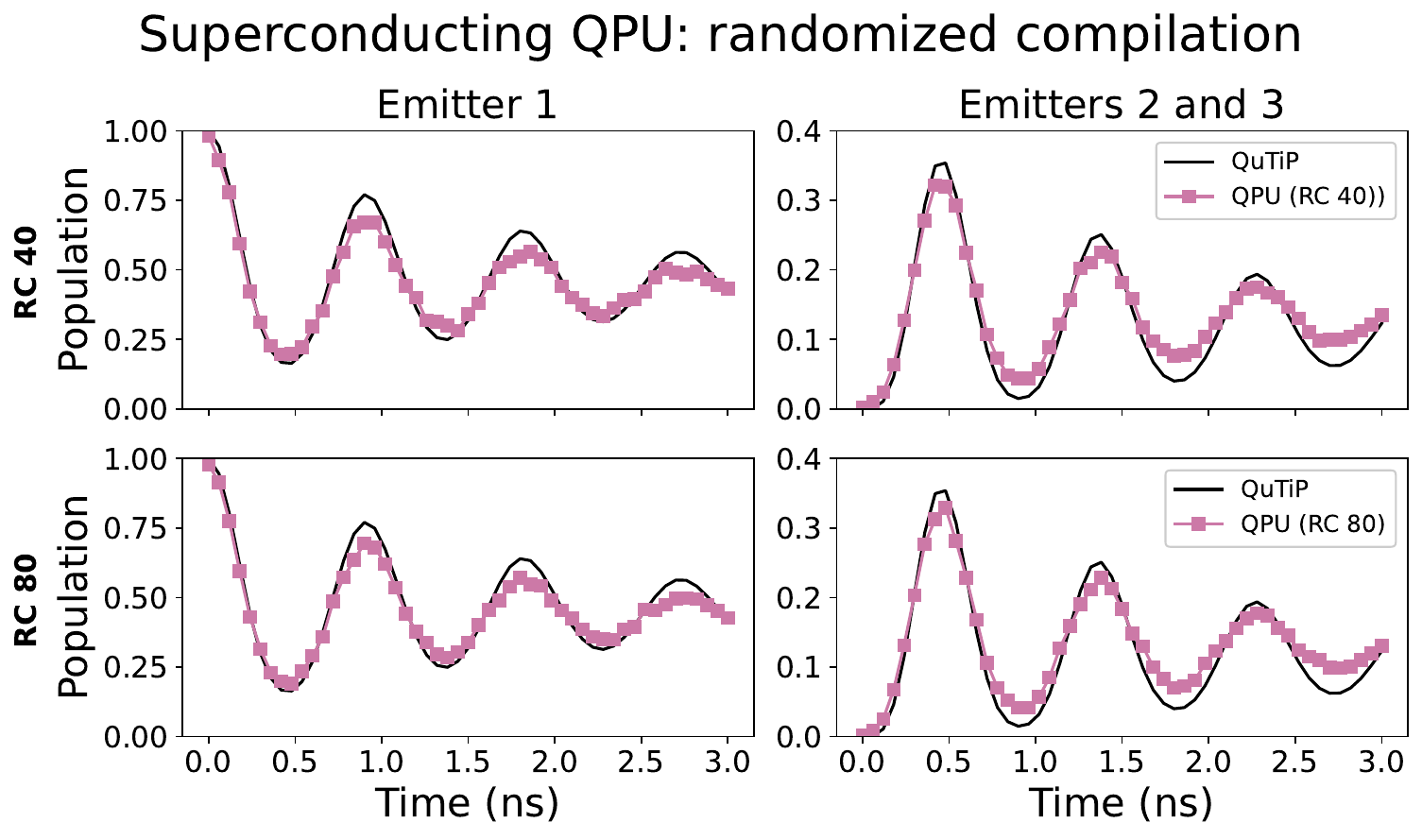}
    \caption{
    \textcolor{red}{
    Results from the AQT superconducting QPU with randomized compilation applied, using 40 (top row) and 80 (bottom row) random compilations.
    Each data point represents a simulation with 20,000 shots; approximately 25\% of shots were discarded in postselection using the method described in Sec.~\ref{sec:postselect}.
    Wall-clock runtime of the simulation was about 17 minutes for 40 randomizations and 24 minutes for 80 randomizations.
    The black line represents the exact solution of the system.
    }
    }
    \label{fig:aqt-rc}
\end{figure}

\textcolor{red}{
We applied randomized compiling (described in Sec.~\ref{sec:rc}) to generate the circuits executed on AQT's superconducting hardware, using 40 and 80 random compilations; the results are presented in Fig.~\ref{fig:aqt-rc}.
Already at 40 randomizations the observed Rabi oscillations are markedly smoother than without error mitigation (Fig.~\ref{fig:aqt-raw}), though increasing to 80 randomizations does not significantly change the results.
}

\subsection{Error mitigation: noiseless output extrapolation (NOX)}
NOX is an error mitigation technique that aims to create a unbiased estimate of what a circuit's output would be if it were operating with zero error \cite{ferracin2022efficiently}.
It achieves this by deliberately increasing the amount of error in the circuit, then extrapolating from these results back to zero error.
To achieve this, each gate cycle $\mathcal{H}$, which is affected by some noise $\mathcal{D}$, is modified so that the noise is selectively amplified: $\mathcal{D}_{\mathcal{H}}\mathcal{H} \rightarrow \mathcal{D}_{\mathcal{H}}^{\alpha} \mathcal{H}$ for some $\alpha > 1$.
The typical noise amplification technique is known as ``identity insertion", which works by replacing $\mathcal{H}$ by $\mathcal{H}(\mathcal{H} \mathcal{H}^{-1})^{\alpha}$. Provided that the $\mathcal{H}$ and its inverse suffer the same noise, and commute with it, identity insertion efficiently amplifies the gate noise without changing the logical operation.

\begin{figure}[htbp]
    \centering
    \includegraphics[width=\columnwidth]{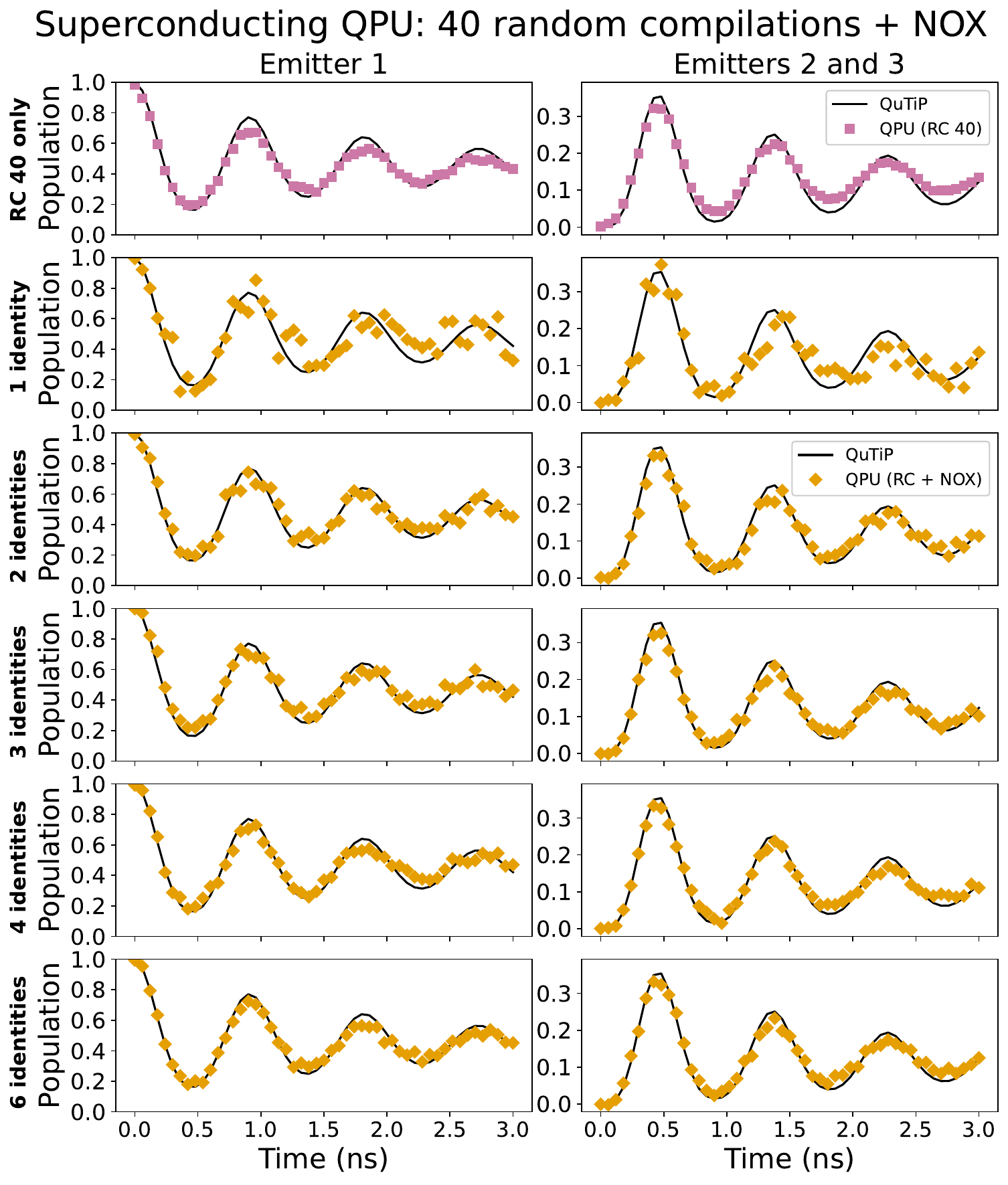}
    \caption{
    \textcolor{red}{
    Results from the AQT superconducting QPU using RC (40 random compilations) and NOX error mitigation with an increasing number of inserted identities.
    Each data point represents a simulation with 20,000 shots; in the NOX data, approximately 8\% of shots were discarded in postselection using the technique described in Sec.~\ref{sec:postselect}.
    Wall-clock runtime of the simulation was about four hours for each iteration of NOX.
    The black line represents the exact solution of the system.
    }
    }
    \label{fig:aqt-rc-nox}
\end{figure}

NOX requires errors to be stochastic in order to accurately estimate the zero-noise condition, making it a natural addition to RC which tailors coherent errors into stochastic ones.
We applied NOX to the simulation circuits generated by random compilation (with 40 randomizations) in order to further suppress noise; the results from executing this version of the algorithm are seen in Fig.~\ref{fig:aqt-rc-nox}.

\begin{figure}[htbp]
    \centering
    \includegraphics[width=\columnwidth]{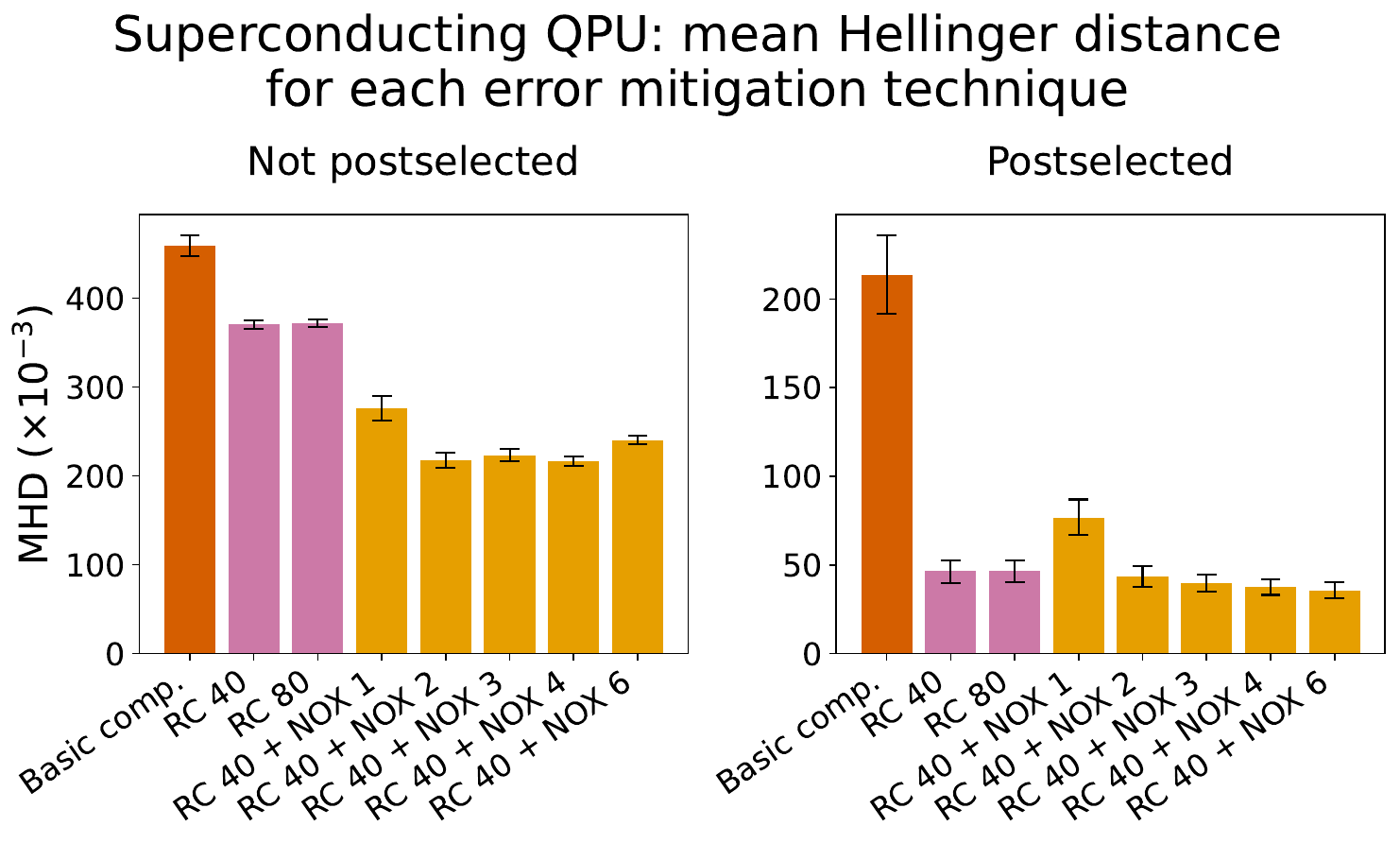}
    \caption{
    \textcolor{red}{
    Mean Hellinger Distance (see Eq.~(\ref{eq:hellinger})) between the simulated distribution and the exact solution obtained with QuTiP for each version of the algorithm executed on the AQT superconducting QPU. 
    Each version of the simulation algorithm propagated the initial state forward by 3 ns in 51 time steps, with 20,000 shots per time step.
    All data has been postselected in the manner described in Sec.~\ref{sec:postselect}.
    Error bars represent 95\% confidence intervals computed via bootstrap resampling with 1,000 replicates.
    }
    }
    \label{fig:aqt-mhd}
\end{figure}

\begin{figure}[htbp]
    \centering
    \includegraphics[width=\columnwidth]{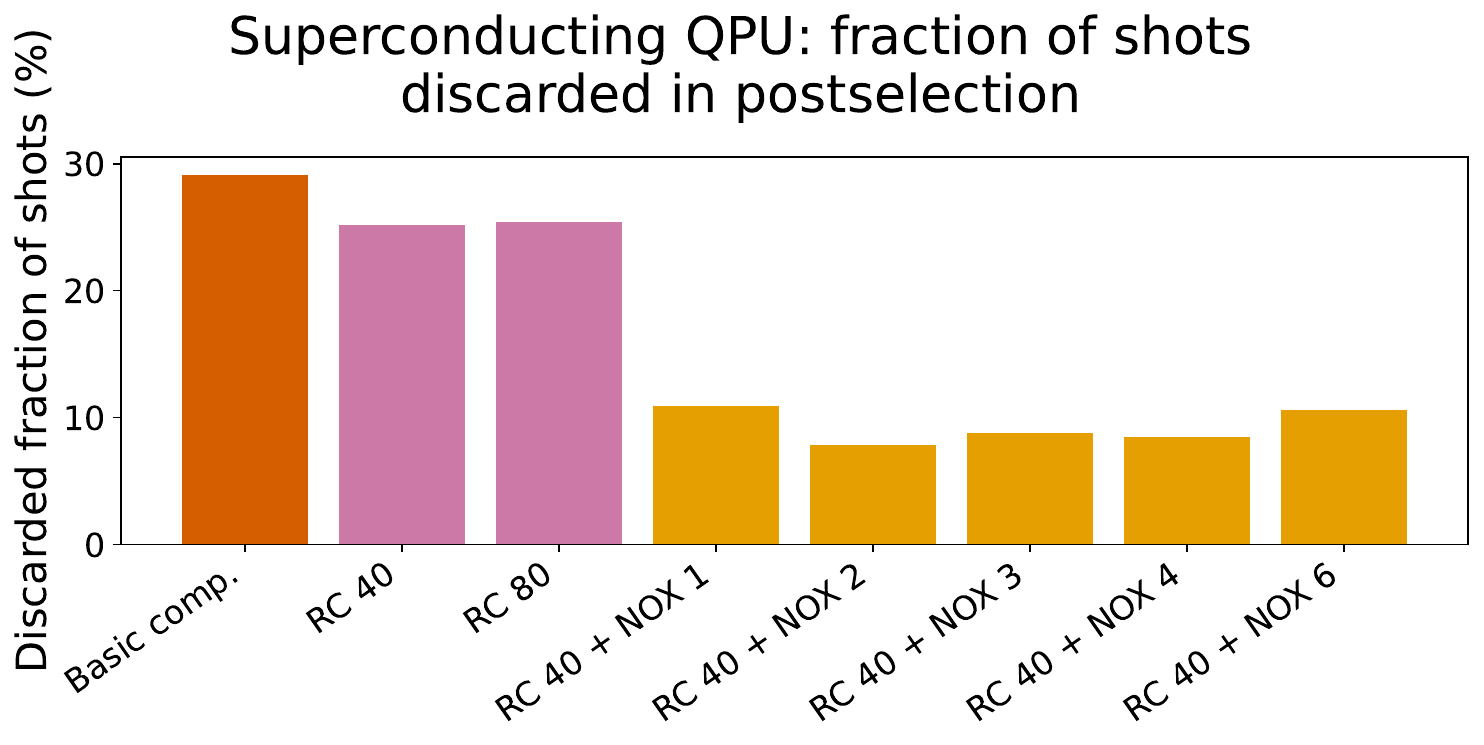}
    \caption{
    \textcolor{red}{
    Fraction of simulation shots discarded by postselection (see Sec.~\ref{sec:postselect}) for each version of the algorithm executed on the QSCOUT trapped ion QPU.
    }
    }
    \label{fig:aqt-discarded}
\end{figure}

\subsection{Analysis}
As described in Section~\ref{sect:hellinger}, we use the \textcolor{red}{Mean Hellinger Distance} to quantify the deviation between the simulation results from the quantum processor and the exact solution obtained numerically with QuTiP.
\textcolor{red}{The MHD for the three methods of error mitigation used on the AQT processor are presented in Fig.~\ref{fig:aqt-mhd}.}
When randomized compiling is applied, coherent errors in the simulation are tailored into stochastic noise whose average better represents the error-free result, curtailing the acute, random deviations from the exact solution which characterize the result without RC.
In general, more random compilations will reduce coherent errors to a greater degree; this effect appears to be saturated already with 40 compilations, as evidenced by the lack of improvement when increasing to 80 random compilations.

When NOX is used, we sample the circuit's output at each level of amplified noise.
The noise amplification needs to be large relative to the sampling error of the original circuit or else the variance of the estimate of the zero noise circuit operation will be large, leading to a larger \textcolor{red}{MHD}. 
We observe a crossover at $\alpha=4$ identities where the \textcolor{red}{MHD} of the zero noise extrapolated estimate falls below that obtained with randomized compiling alone.

\textcolor{red}{We can also see the strength of NOX in mitigating error in Fig.~\ref{fig:aqt-discarded} which shows the fraction of shots across the entire simulation discarded in the postselection process.
This fraction remains above 20\% even when RC is applied, but NOX reduces it below 10\%.}

\begin{figure}[htbp]
    \centering
    \includegraphics[width=\columnwidth]{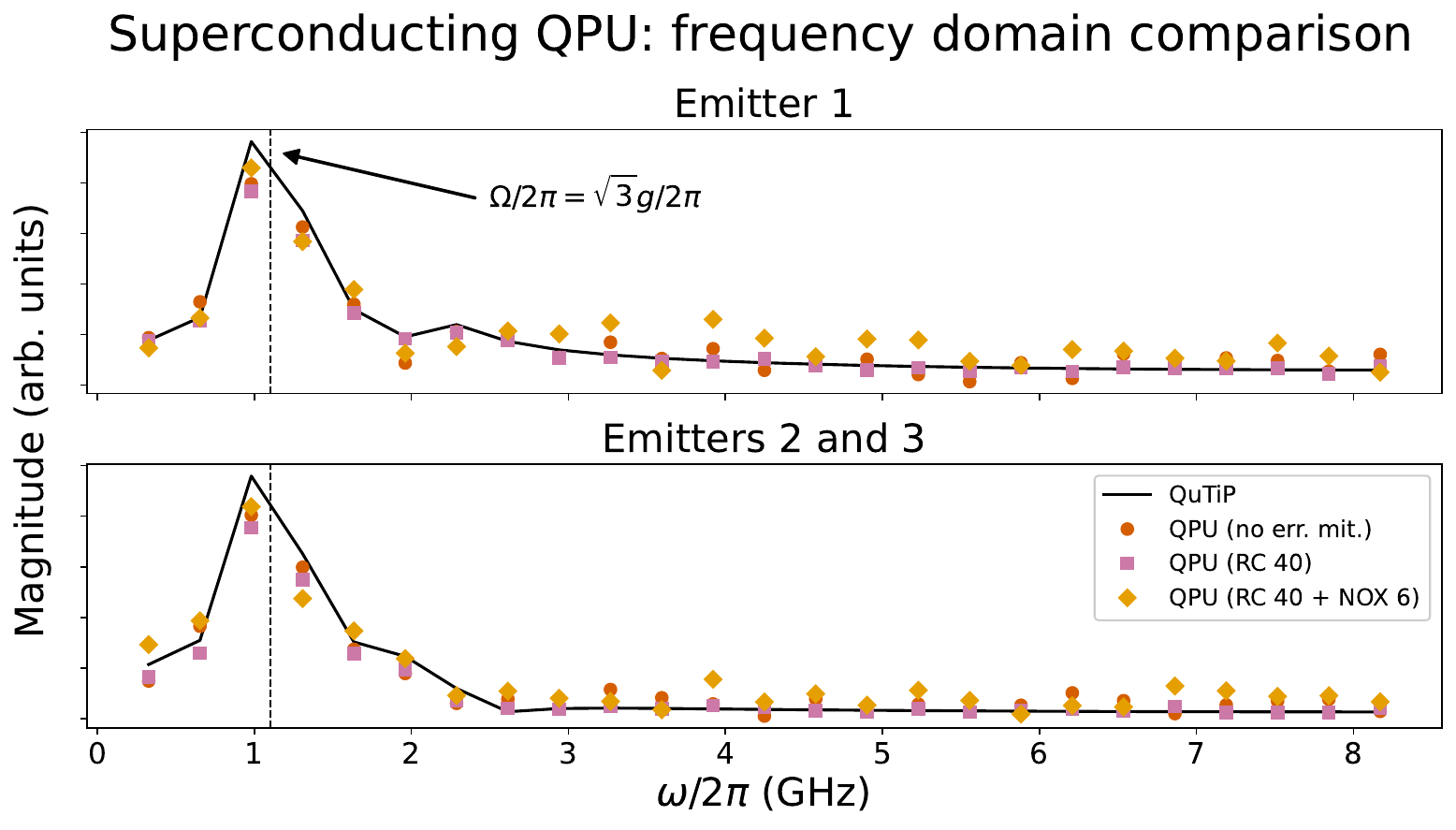}
    \caption{
    \textcolor{red}{
    Fourier transforms of the simulation results from the AQT superconducting QPU.
    The mean population has been subtracted, to eliminate the zero-frequency component.
    The underlying population data has been postselected according to the procedure described in Sec.~\ref{sec:postselect}.
    The dashed line indicates the expected Rabi frequency for the simulated system.
    The black line indicates the Fourier transform of the numerical solution of the system obtained with the Python package QuTiP.
    }
    }
    \label{fig:aqt-fourier}
\end{figure}

As in Sec.~\ref{sect:hellinger}, we compare the Fourier transforms of the simulation results with those of the numerical solution obtained with QuTiP.
These Fourier transforms are presented in Fig.~\ref{fig:aqt-fourier} for all three versions of the algorithm executed on the AQT superconducting processor.
As with the results from the QSCOUT processor, we find that the quantum simulations accurately obtain the Rabi rate of the system even without error mitigation.

\section{Discussion}
These results show the potential for today’s NISQ-era quantum computers to \textcolor{red}{accurately} simulate open, many-body cavity QED physics, and the vital role that error mitigation plays in reducing noise in these simulations.
\textcolor{red}{The contrast between the two platforms used in this work highlights important trade-offs in quantum computing architectures.
Each platform offers unique advantages: trapped ions provide native all-to-all connectivity ideal for CQED simulations, while superconducting qubits offer faster gate times and greater system stability.
The nature of the noise affecting each system is also significantly different, as a consequence of arising from different physical mechanisms.
As we have seen, these distinct architectural and noise characteristics have strong implications for which error mitigation techniques are most effective.}

We have shown that error mitigation techniques such as RC, NOX, mirror SWAPs and optimized compilation techniques, selected on the basis of the platform at hand, can be effective at suppressing noise enough to obtain \textcolor{red}{accurate} results in the simulation of open CQED systems.
\textcolor{red}{The effectiveness of different error mitigation strategies depends strongly on the characteristic noise of each platform.
Superconducting qubits primarily suffer from coherent errors that are relatively stable over time, arising from miscalibrated gate pulses or slow variation in flux voltages.
This makes them well-suited to randomized compiling, which converts these coherent errors into stochastic ones.
Such stochastic errors are then effectively suppressed by NOX, since the errors present in a given gate cycle will also be present in the additional copies of that gate (and its inverse) used for identity insertion \cite{ferracin2022efficiently}.
The stability of superconducting systems and their fast gate times allows them to support the relatively deep circuits generated by RC and to take a larger number of shots per simulation run, which improves statistical precision.}

\textcolor{red}{By contrast, ion traps exhibit noise of a more strongly stochastic nature due to trap heating and fluctuations in trap frequency.
This fundamental difference in noise character makes RC less effective for trapped ions, and NOX particularly challenging to implement \cite{maupin2023error}.
On the other hand, our postselection technique is particularly powerful in this situation, even in the absence of other error mitigation techniques.
This can be seen by comparing the most basic version of the algorithm on each platform: postselection improves the MHD by roughly a factor of 10 for the hand-compiled version of the algorithm on the trapped ion platform, but only a factor of about 2 for the basic compilation of the algorithm on the superconducting platform.
Ion trap processors also benefit from optimization strategies not focused on the statistics of the noise.
Compilation with variable-angle ZZ($\theta$) entangling gates and SWAP mirroring both reduced the MHD of the simulations, both with and without postselection applied.
Indeed, these techniques are particularly appropriate for our algorithm, which contains many controlled rotations and thus many optimization opportunities for both compilation techniques to capitalize upon.}

\textcolor{red}{While the simulations presented in this work represent small systems which are a special case of the open Tavis-Cummings model, improved simulation algorithms and larger quantum computers could enable simulation of more general open quantum optical systems beyond the scale easily simulated on classical computers.}
\textcolor{red}{The algorithm we have applied relies on a relatively simple mapping between elements of an open CQED system and computational qubits, and only applies in the special case of identical, lossless quantum emitters identically coupled to a lossy cavity.}
\textcolor{red}{Most existing quantum algorithms for simulating open quantum systems are far too demanding in terms of qubit count to be feasible on today's NISQ machines.}
\textcolor{red}{Yet, these algorithms are also highly general, and there may be opportunities for cost savings by considering the aspects of the problem unique to CQED.}

Many-body open CQED holds great promise as a basis for future quantum technologies.
Subradiant states exhibited by groups of inhomogeneous emitters coupled to a cavity may form the basis of quantum optical memories \cite{Lei2023}.
Engineered of dissipation in these systems could be applied to quantum state preparation \cite{PhysRevLett.106.090502}.
Many of these phenomena are challenging to study experimentally, and can be costly to simulate classically above small scales.

By applying emerging techniques of quantum simulation to open CQED systems, we can hope to achieve a virtuous cycle of technological development, in which better simulations help design better quantum devices, which in turn enable better simulations.
However, starting this engine is difficult, as current quantum computers are generally too noisy to support simulations beyond what is achievable with classical computers.
Until and unless large, error-corrected quantum computers become available, error mitigation will be vital to obtaining \textcolor{red}{accurate} simulation results.
This work stands as a platform-specific guide to the best use of error mitigation in quantum simulation of open cavity quantum electrodynamical systems.

\section{Acknowledgments}
Authors acknowledge support by Noyce Foundation, National Science Foundation CAREER program (Award 2047564), the UC Multicampus Research Programs and Initiatives of the University of California (Grant Number M23PL5936), Google Research Scholar Fellowship, and the Pauli Center for Theoretical Study. 
This material was funded in part by the U.S. Department of Energy, Office of Science, Office of Advanced Scientific Computing Research Quantum Testbed Program under contracts DE-AC02-05CH11231 and DE-NA0003525.
Sandia National Laboratories is a multi-mission laboratory managed and operated by National Technology and Engineering Solutions of Sandia, LLC, a wholly owned subsidiary of Honeywell International Inc., for the U.S. Department of Energy’s National Nuclear Security Administration under contract DE-NA0003525.
This paper describes objective technical results and analysis.
Any subjective views or opinions that might be expressed in the paper do not necessarily represent the views of the U.S. Department of Energy or the United States Government.

\section{Data availability}
All data from the quantum simulations, as well as the code used to solve the model with QuTiP and produce the figures in this work can be found at \url{github.com/radulaski/CQED_quantum_simulation}.

\nocite{*}

\bibliographystyle{unsrt}
\bibliography{main}

\appendix
\section{QSCOUT gate fidelities \label{app:qscout-fidelity}}
\textcolor{red}{
The data from QSCOUT was collected in four runs with 500 shots each.
In order to minimize the effect of system drift on the results, the four versions of the circuit were interleaved within each run.
Tables~\ref{tab:qscout-1}-\ref{tab:qscout-4} give the fidelities of the 2-qubit bare MS gate measured during each run, for each pair of qubits used in the circuit.
Single qubit gate fidelities for these runs were 99.3\%.
}

\renewcommand{\arraystretch}{1.3}
\begin{table}[h]   
    \begin{tabular}{|c|c|c|}
    \hline
        Gate Pair & Connectivity & Fidelity Upper Bound \\
    \hline
    \{q[0], q[1]\} & nearest neighbor & $0.984_{-0.009}^{+0.008}$ \\
    \hline
    \{q[0], q[2]\} & nearest neighbor & $0.993_{-0.008}^{+0.006}$ \\
    \hline
    \{q[0], q[3]\} & next-nearest neighbor & $0.976_{-0.010}^{+0.009}$ \\
    \hline
    \{q[1], q[3]\} & nearest neighbor & $0.992_{-0.008}^{+0.006}$ \\
    \hline
    \{q[2], q[3]\} & outer & $0.985_{-0.009}^{+0.008}$ \\
    \hline
    \end{tabular}
\caption{Bare MS entangling gate fidelities from QSCOUT run 0.}
\label{tab:qscout-1}
\end{table}

\begin{table}[h]   
    \begin{tabular}{|c|c|c|}
    \hline
        Gate Pair & Connectivity & Fidelity Upper Bound \\
    \hline
    \{q[0], q[1]\} & nearest neighbor & $0.975_{-0.010}^{+0.009}$ \\
    \hline
    \{q[0], q[2]\} & nearest neighbor & $0.983_{-0.009}^{+0.008}$ \\
    \hline
    \{q[0], q[3]\} & next-nearest neighbor & $0.983_{-0.010}^{+0.008}$ \\
    \hline
    \{q[1], q[3]\} & nearest neighbor & $0.990_{-0.009}^{+0.007}$ \\
    \hline
    \{q[2], q[3]\} & outer & $0.978_{-0.010}^{+0.008}$ \\
    \hline
    \end{tabular}
\caption{Bare MS entangling gate fidelities from QSCOUT run 1.}
\label{tab:qscout-2}
\end{table}

\begin{table}[h]   
    \begin{tabular}{|c|c|c|}
    \hline
        Gate Pair & Connectivity & Fidelity Upper Bound \\
    \hline
    \{q[0], q[1]\} & nearest neighbor & $0.988_{-0.008}^{+0.006}$ \\
    \hline
    \{q[0], q[2]\} & nearest neighbor & $0.990_{-0.009}^{+0.006}$ \\
    \hline
    \{q[0], q[3]\} & next-nearest neighbor & $0.984_{-0.009}^{+0.008}$ \\
    \hline
    \{q[1], q[3]\} & nearest neighbor & $0.988_{-0.008}^{+0.006}$ \\
    \hline
    \{q[2], q[3]\} & outer & $0.981_{-0.010}^{+0.008}$ \\
    \hline
    \end{tabular}
\caption{Bare MS entangling gate fidelities from QSCOUT run 2.}
\label{tab:qscout-3}
\end{table}

\begin{table}[h]   
    \begin{tabular}{|c|c|c|}
    \hline
        Gate Pair & Connectivity & Fidelity Upper Bound \\
    \hline
    \{q[0], q[1]\} & nearest neighbor & $0.995_{-0.007}^{+0.005}$ \\
    \hline
    \{q[0], q[2]\} & nearest neighbor & $0.988_{-0.009}^{+0.007}$ \\
    \hline
    \{q[0], q[3]\} & next-nearest neighbor & $0.988_{-0.009}^{+0.007}$ \\
    \hline
    \{q[1], q[3]\} & nearest neighbor & $0.994_{-0.007}^{+0.005}$ \\
    \hline
    \{q[2], q[3]\} & outer & $0.985_{-0.009}^{+0.007}$ \\
    \hline
    \end{tabular}
\caption{Bare MS entangling gate fidelities from QSCOUT run 3.}
\label{tab:qscout-4}
\end{table}

\end{document}